\documentclass[a4paper,fleqn]{cas-dc}

\usepackage[numbers]{natbib}

\def\tsc#1{\csdef{#1}{\textsc{\lowercase{#1}}\xspace}}
\tsc{WGM}
\tsc{QE}
\tsc{EP}
\tsc{PMS}
\tsc{BEC}
\tsc{DE}

\begin{document}
\let\WriteBookmarks\relax
\def\floatpagepagefraction{1}
\def\textpagefraction{.001}
\shorttitle{FNO-CCSDT: A computational study}
\shortauthors{Manisha et~al.}

\title [mode = title]{Frozen Natural Orbitals based Equation-of-motion coupled-cluster singles, doubles and triples for Ionized, Double-Ionized, Electron Attached and Two-Electron Attached states }                      



\author[]{Manisha}[type=editor,
                         bioid=1]

\credit{Methodology, Software, Validation and Formal analysis, Writing - Original draft preparation}

\affiliation[]{organization={Department of Chemistry, Birla Institute of Technology and Science},
                city={Pilani},
                postcode={333031}, 
                state={Rajasthan},
                country={India}}

                \author[]{Prashant Uday Manohar}[type=editor,
                         bioid=1,
                 orcid=0000-0001-9788-4460]
\cormark[1]
\ead{pumanohar@pilani.bits-pilani.ac.in}

\credit{Conceptualization of this study, Writing – Review and Editing, Supervision}


\cortext[cor1]{Corresponding author}

\begin{abstract}
In this work, we present frozen natural orbital (FNO) based implementations of equation-of-motion (EOM) coupled-cluster (CC) with singles, doubles, and triples (SDT) for ionization potential (IP), double ionization potential (DIP), electron attachment (EA), and double electron attachment (DEA) variants. For EOM-CC with singles and doubles (SD), the FNO approach has already been studied by Krylov and co-workers for IP variant and for spin-flipping and spin-conserving excited states (respectively, the SF and EE variants) for both total energies and energy-gaps. Recently, we presented FNO-CCSDT performance for ground state energies of molecules, triplet-singlet gaps and for numerical estimation force constants of some diatomic molecules.
Now we present our study on performance of IP, DIP, EA and DEA variants of FNO-EOM-CCSDT in computing total-energies, and for target-reference and target-target energy-gaps. Following earlier studies by us and by Krylov and co-workers, we also present the XFNO-EOM-CCSDT approach for these variants and examine its performance for total energies and energy-gaps.



\end{abstract}




\maketitle

\section{Introduction}
Accurately capturing electron correlation \cite{helgaker2013molecular} is central to making reliable predictions in quantum chemistry. However, methods that rigorously account for correlation—such as CC\cite{kummel2003biography,vcivzek1966correlation,sinanouglu1970three,bartlett2007coupled} and many-body perturbation theory (MBPT)—come with steep computational costs. As molecular systems grow in size, the scaling quickly becomes a barrier. For example, second-order Møller–Plesset perturbation theory (MP2)\cite{moller1934note} scales as $O(N^5)$, CCSD \cite{schutz2001low,purvis1982full} as $O(N^6)$, CCSD(T) \cite{Lee1995, BARTLETT1990513} as $O(N^7)$, and CCSDT \cite{noga1987full} as $O(N^8)$, where $N$ indicates computational size of the system which is dependent on the number of electrons and the basis functions. This rapid growth in cost limits the application of these high-accuracy methods to small- or medium-sized systems. Variety of cost-reduction strategies can overcome these challenges to some extent, thereby widening the applicability of these methods. For example, the techniques such as Cholesky Decomposition (CD) \cite{vysotskiy2011accurate,kumar2017resolution, kumar2019analytical, epifanovsky2013general, 10.1063/1.1578621} and Resolution of Identity (RI) \cite{vogt2008accelerating,kumar2017resolution, epifanovsky2013general, https://doi.org/10.1002/qua.560120408} are density-fitting approaches in which the electron-repulsion integrals (ERI)s are approximated as sum-over-products of one-electron densities resulting in reduced storage and memory requirements. These approaches can significantly speed-up CCSD and lower-order perturbation methods (MP2, MP3, MP4), in which the four-index quantities (ERIs, $T_2$ amplitudes, etc) are the costliest from computational perspective. For CCSDT and EOM-CCSDT, the rate-determining quantities being the six-index quantities ($T_3$, $R_3$, $L_3$ amplitudes), the cost-cutting in ERI computation would hardly affect the computational speed, though these techniques have already been reported.
 Use of FNO \cite{taube2008frozen, taube2005frozen,10.1063/1.5138643,10.1063/1.3276630} is another attractive cost-effective approach due to its ability to shrink the virtual orbital space in post-Hartree–Fock methods with minimal loss of accuracy.
 The foundation of the FNO approach lies in Natural Orbitals (NOs), which are defined as the eigenfunctions of the one-particle reduced density matrix (1-RDM) derived from an N-electron wave function. The corresponding eigenvalues—known as natural occupation numbers—indicate how much each orbital contributes to electron correlation. In Hartree–Fock theory, these occupation numbers are either 1 or 0, but in correlated methods, they become fractional, offering valuable insight into the relevance of each orbital. This property makes NOs a compact and physically meaningful basis for correlation    methods. Orbitals with larger occupation numbers typically contribute more to the total correlation energy. By retaining only those with significant occupation and discarding the rest, one can achieve substantial computational savings without sacrificing much accuracy—this is the essence of the FNO technique. The idea traces back to Löwdin’s \cite{lowdin1955quantum} work in the 1950s, where he demonstrated that expanding the wave function in a natural orbital basis improved the convergence of configuration interaction (CI) calculations. Barr and Davidson \cite{barr1970nature} later formalized the FNO approach by rotating only the virtual orbitals into the NO basis, while keeping the occupied orbitals fixed at their Hartree–Fock values. This formulation ensures that the reference energy and correlation energy remain unchanged under the transformation, making it a practical and elegant solution. Since then, FNOs have been widely used across a range of correlated methods. For instance, Shavitt \cite{shavitt1976comparison} and co-workers showed that using NOs significantly accelerated CI convergence for small molecules like water. FNOs also play an important role in multi-reference methods, where they help define active spaces based on occupation number thresholds. Practically, generating NOs is quite efficient. A common strategy is to compute them from a lower-level method like MP2 and use them in a more accurate method such as CCSD or CASSCF. This two-step approach allows the virtual space to be meaningfully reduced without compromising the core physics of the system. MP2-derived NOs\cite{jensen1988second} are also often used to initialize CASSCF \cite{abrams2003comparison}calculations, helping avoid the need for expensive orbital optimization. In coupled-cluster and perturbative methods, FNOs have been particularly influential. Benchmark studies by Bartlett and co-workers demonstrated that truncating up to 60\% of the virtual orbitals in CCSD(T) calculations led to only minor errors in bond lengths—typically within 0.2–0.3 pm—when using large basis sets like cc-pVTZ. Krylov and her group later incorporated the FNO framework into EOM-CCSD methods for IP, and extended it to open-shell references via the OSFNO approach\cite{10.1063/1.5138643}, enabling accurate and efficient treatment of spin-flip excited states. Building on this foundation, our group has recently applied the FNO method together with single-precision arithmetic to full CCSDT calculations. The observed energy deviations remained within 0.5–1 kcal/mol—well within the accepted error margin for CCSDT—highlighting the effectiveness and robustness of this combined cost-saving strategy, even at such a high level of theory. Extending this work further, we have developed FNO-based implementations for both IP-EOM-CCSDT, DIP-EOM-CCSDT, EA-EOM-CCSDT, and DEA-EOM-CCSDT methods, supporting both single and double precision algorithms. In the sections that follow, we briefly introduce the theoretical background of the FNO method and then present detailed results for the FNO-IP-EOM-CCSDT, FNO-DIP-EOM-CCSDT, FNO-EA-EOM-CCSDT, and FNO-DEA-EOM-CCSDT approaches. We conclude with a discussion on the performance and accuracy of these methods, and their potential to enable high-level quantum chemical calculations for larger, more complex molecular systems. 
 
\subsection{The EOM-CCSDT method}
The FNO scheme is well known for ground state CCSDT method.
We are extending this approach to electronically ionized, double ionized, electron attached and double electron attached states . The IP-EOM-CC describes open-shell doublet states as “ionized” states derived from a well-behaved closed-shell reference wave function whereas the double DIP-EOM-CC will behave as  close-shell double ionized state similarly for the electron attached and double electron attached states can be defined. The wavefunction for the IP, DIP, EA, DEA EOM-CCSDT methods is described as follows.

\begin{equation}
|\Psi_a^{(N-m)}\rangle ={\hat R}_a^{(-m)}|\Psi_{0}^{(N)}\rangle  \label{eq:eomform}
\end{equation}
\begin{equation}
|\Psi_a^{(N+m)}\rangle ={\hat R}_a^{(+m)}|\Psi_{0}^{(N)}\rangle  \label{eq:eomform1}
\end{equation}
where ${\hat R}_{a}^{(-m)}$ is the linear operator and $m=1$ for IP and $m=2$ for DIP states similarly ${\hat R}_{a}^{(+m)}$  is the linear operator and  $m=1$ for EA and $m=2$ for DEA states.
\begin{equation}
{\hat R}_a^{(-m)}=\sum\limits_{n=m}^{M_R} {\hat R}_{a;nh,(n-m)p}  \label{eq:eomreq}
\end{equation}
\begin{equation}
{\hat R}_a^{(+m)}=\sum\limits_{n=m}^{M_R} {\hat R}_{a;nh,(n+m)p}  \label{eq:eomreq1}
\end{equation}
where ${\hat R}_{a;nh,(n-m)p}$ representing $n$ hole–$n-m$-particle components, and removes $m$ electron(s) whereas ${\hat R}_{a;nh,(n+m)p}$ representing $n$ hole $(n+m)p$ components, and add $m$ elctron(s) from the CC ground state which described as:
\begin{equation}
|\Psi_{0}^{(N)}\rangle =e^{T}|\Phi_{0}\rangle  \label{eq:ccwfn}
\end{equation}
The operator $\hat T$ is an excitation operator satisfying the reference-state CC equations
\begin{eqnarray}
0=\langle\Phi^{n}|\Bar{H}|\Phi_{0} \rangle
=\langle\Phi^{n}|e^{-\hat T}{\hat H}e^{\hat T}|\Phi_0 \rangle \label{eq:cc_t} \\
E=\langle\Phi_{0}|\Bar{H}|\Phi_{0} \rangle
=\langle\Phi_{0}|e^{-\hat T}{\hat H}e^{\hat T}|\Phi_0 \rangle \label{eq:cc_e} \\
\end{eqnarray}
Here, the $\Phi^n$ represent $n$-tuple excited Slater determinants with respect to the reference determinant, $\Phi_0$. In FNO-based calculations, both the cluster operator $\hat{T}$ and the excitation operators used to construct the EOM wave functions are defined within the truncated (active) virtual orbital space. The EOM wave functions themselves are obtained as eigenstates of the similarity-transformed Hamiltonian,$\Bar H = e^{-\hat{T}}\Hat{H}e^{\hat{T}}$.
In IP-EOM-CCSD, DIP-EOM-CCSD  the equation-of-motion operators coupled cluster is limited to singles and doubles—meaning we account for excitations like (2h1p)  for IP whereas (3h1p) for DIP. To boost accuracy, especially for systems where subtle effects matter, we can include triple excitations which results to (3h2p) for ionized state and (4h2p) for double ionized state.
Once we’ve generated the FNOs and trimmed down the virtual space, all calculations are carried out in this reduced orbital set—often called the active space. Truncating the virtual orbitals changes two things: the structure of the similarity-transformed Hamiltonian ($\bar{H} = e^{-\hat{T}} \hat{H} e^{\hat{T}}$), and the space in which we solve for the target states using the EOM operator.
Ideally, the eigenvalues of $\bar{H}$ shouldn’t depend on how we treat correlation. But when we truncate the virtual space, we risk losing important contributions where excited states may involve orbitals not crucial to the ground state and thus omitted.

EOM-IP and DIP, however, is more robust. Since it mainly involves electron removal, it relies less on those excluded orbitals. While truncation may slightly affect correlation and relaxation via 3h2p and 4h2p terms, our results show the accuracy remains largely intact—making FNO a smart, cost-effective choice for EOM-IP and DIP calculations.

\subsection{FNO approximation}
 The frozen natural orbital (FNO) approach offers an efficient strategy to reduce the virtual orbital space in correlated wavefunction methods, significantly lowering computational cost with minimal impact on accuracy. Bartlett and co-workers extensively investigated this technique, including its application to CCSD(T) \cite{taube2005frozen,taube2008frozen}. In the context of excited-state methods, Krylov and co-workers implemented the FNO-IP-EOM-CCSD approach using a closed-shell reference and later extended it to open-shell systems through the open-shell FNO (OSFNO) framework, enabling its application to spin-flip (SF) formulations. 
 Here, we briefly outline the general formulation of FNO. The key idea is to transform the virtual Hartree–Fock (HF) orbitals into a new set of natural orbitals (NOs), which are then selectively truncated based on their relative importance to electron correlation. This transformation is achieved using the second-order Møller–Plesset (MP2) approximation to construct a virtual–virtual block of the unrelaxed one-particle density matrix (DM). 

The first-order correction to the HF wavefunction is expressed as a linear combination of doubly excited determinants, whose amplitudes are evaluated at the MP2 level using:
\begin{eqnarray}
 t_{ij}^{ab}(\text{MP2}) = \frac{\langle ab||ij\rangle}{\epsilon_{ij}^{ab}}, \label{eq:t2mp2}
\end{eqnarray}
where $\langle ab||ij\rangle$ are antisymmetrized two-electron Coulomb integrals, and the energy denominator is defined as:
\begin{eqnarray}
 \epsilon_{ij}^{ab} = f_{ii} + f_{jj} - f_{aa} - f_{bb}, \label{eq:enedenom}
\end{eqnarray}
with $f_{pp}$ representing the orbital energies of the canonical MOs.

Using these MP2 amplitudes, we construct the virtual–virtual block of the second-order density matrix:
\begin{eqnarray}
 D^{(2)}_{ab} = \frac{1}{2} \sum\limits_{cij} \frac{\langle cb||ij\rangle \langle ij||ca\rangle}{\epsilon_{ij}^{cb} \epsilon_{ij}^{ca}}, \label{eq:mp2dm}
\end{eqnarray}
which is then diagonalized to obtain the NOs. The eigenvectors define the transformation from canonical virtual orbitals to natural orbitals, and the eigenvalues correspond to the NO occupation numbers. These NOs are then sorted in descending order of occupation. Those with the highest occupation are retained as the active virtual space for subsequent correlated calculations, while orbitals with low occupation—typically the least correlated ones—are frozen.

Two different truncation schemes may be employed at this stage, depending on the level of accuracy desired. Once the active NOs are selected, they are optionally transformed into a semicanonical form by block-diagonalizing the Fock matrix within the virtual space. This transformation helps accelerate convergence in coupled-cluster iterations and ensures a block-diagonal structure of the Fock operator consisting of (i) frozen core orbitals, (ii) active occupied orbitals, (iii) semicanonical active virtual NOs, and (iv) frozen virtual NOs. Importantly, these transformations do not affect the separation between occupied and virtual spaces—the off-diagonal virtual–occupied blocks of the Fock matrix remain zero.

In energy calculations, the frozen virtual NOs are simply excluded from the correlation treatment. However, in analytic gradients and property evaluations that involve orbital response, the contribution of these inactive orbitals must be explicitly included, akin to the treatment of frozen core and frozen virtual orbitals in standard response theory.

\section{Computational Details}
 All of the multi-threaded parallel calculations were performed using a development version of the Q-Chem software package. The specific basis sets used for each calculation are listed in the respective tables, and the optimized molecular geometries can be found in the Supplementary Information (SI). For the IP, DIP, EA, and DEA calculations within the XFNO-EOM-CCSDT framework, we used three different occupation-based truncation thresholds—99.9\%, 99.75\%, and 99.5\%—as for the extrapolation of FNO approach to improve accuracy. For the ease of discusion, we refer to the standard EOM-CCSDT results as $M_1$, the FNO-EOM-CCSDT results computed with the 99.9\% threshold as $M_2$, and the extrapolated XFNO-EOM-CCSDT values as  $M_3$. Hereafter, TEs refer to the total energies, whereas MAE refers to the maximum absolute error. The detailed numerical data used to generate the  $M_3$ extrapolations are provided in the Supplementary Information.

\section{Results and Discussion}
\protect\label{results_discussion}
 \subsection{FNO- and XFNO-EOM-CCSDT methods for IP and DIP states}

\begin{table}[h!]
\centering
\renewcommand{\arraystretch}{1.5} 
\setlength{\tabcolsep}{4pt} 
\tiny 
\caption{\small EOM-CCSDT total energies (in hartrees) and relative errors in FNO-CCSDT ($M_2$) and XFNO-EOM-CCSDT  ($M_3$) energies for IP and DIP states (in millihartrees).}
\protect\label{tab:XFNO}
\begin{tabular}{*{7 }{>{\tiny \arraybackslash}c}}
\hline
 &\tiny Molecules& \tiny Basis &  \tiny $E_{\text{$M_1$}}$   &  \tiny $\delta E_{\text{$M_2$}}$ &  \tiny $  \delta E_{\text{$M_3$}}$   \\   
\hline
\hline
\renewcommand{\arraystretch}{1.2} 
&$C_2H_2$	(X$^2\Pi_u$)&cc-pVTZ& -76.76790&		1.031&	0.037\\
&$HCl$  (X$^1\Sigma^+$) &cc-pVTZ&	-459.87546&		0.469&	-0.003\\
&$SC$ (X$^1\Sigma^+$) &cc-pVTZ&	-435.27782&		0.969&	-0.104\\
&$CO$ (X$^2\Sigma^+$) &cc-pVTZ& -112.64495&	 	1.035&	-0.213\\
\textbf{IP}&$NO$ (X$^2\Pi$)&cc-pVTZ&	-129.36731&	 	1.955&	0.278\\
&$NB_2^+$ (X$^2\Sigma^+_g$)&cc-pVDZ&-104.02082&	 	0.506&	0.192\\
&$N_2$ (X$^2\Sigma^+_g$)&cc-pVTZ &	-108.80600&	 	1.281 &  -0.464\\
&$C_2H_4$ (X$^2B_{3u}$)&cc-pVDZ&  -77.97551&	 	1.414&	1.002\\
 \hline
 &$CO$ ($^1\Sigma^+$) &aug-cc-pVTZ&-111.64008&1.286&0.323\\
&$C_2H_2$ ($^1\Delta_g$)&aug-cc-pVTZ&-75.96576&0.856&0.167\\
&$C_2H_2$ ($^1\Pi_u$)&aug-cc-pVTZ& -75.96576&0.857&0.175\\
&$HCHO$ ($^1A_1$)&aug-cc-pVDZ&-113.03592&0.862&-0.093\\
\textbf{DIP}&$HCHO$ ($^1A_2$)&aug-cc-pVDZ&-112.89773&0.812&-0.296\\
&$C_2H_4$ ($^1A_g$)&aug-cc-pVDZ&-77.24125&0.507&-1.244\\
&$C_2H_4$($^1A_g$)&aug-cc-pVDZ&-77.18424&0.613&-0.604\\
&$C_2H_4$ ($^1B_g$)&aug-cc-pVDZ&-77.09388&0.607&-0.371\\
&$H_2O$ ($^1A_1$)&aug-cc-pVTZ&-74.82057&0.444&-0.092\\
&$H_2O$  ($^1B_1$)&aug-cc-pVTZ&-74.76842&0.403&-0.302\\
\hline

\end{tabular}
\end{table}

 Table \ref{tab:XFNO} presents the $M_1$ TEs (in hartrees)  and the TE errors (in millihartrees) of $M_2$ and $M_3$ relative to $M_1$. $M_2$ systematically overestimates the TEs compared to the $M_1$ benchmarks, for IP and DIP variants, with the MAE of, respectively, 1.4 and 1.3 millihartrees.
 Although the trends in $M_3$ TEs are less systematic, the relative absolute errors are smaller than the corresponding $M_2$ ones. For IP and DIP variants are, the $M_3$ MAEs for TEs are, respectively, 1.0 and 1.2 millihartrees.
 These trends are very similar to the ones observed in our earlier work \cite{FNOCCSDT} on FNO/XFNO-CCSDT TEs.

\begin{table}[h!]
\centering
\renewcommand{\arraystretch}{1.5} 
\setlength{\tabcolsep}{4pt} 
\tiny 
\caption{\small EOM-CCSDT energies and relative errors in $M_2$ and XFNO-EOM-CCSDT energies for IP states(in electron-volts).}
\protect\label{tab:XFNO1}
\begin{tabular}{*{7}{>{\tiny \arraybackslash}c}}
\hline
 \tiny Molecules& \tiny Basis &  \tiny $E_{\text{$M_1$}}$   &  \tiny $\delta E_{\text{$M_2$}}$ &  \tiny $  \delta E_{\text{$M_3$}}$ & $Exp$   \\  
 \\ 
\hline
\hline
\renewcommand{\arraystretch}{1.2} 
$C_2H_2$ (X$^2\Pi_u$)&cc-pVTZ&11.420 &0.001& 0.003&11.40\\
$HCl$  (X$^1\Sigma^+$)&cc-pVTZ&	12.576 &0.002&-0.001&12.75\\
$SC$  (X$^1\Sigma^+$)&cc-pVTZ&	11.306 &0.003&-0.0004& 11.33\\
$CO$  (X$^2\Sigma^+$)&cc-pVTZ&  13.895&0.003&0.0004&14.01\\
$NO$    (X$^2\Pi$)&cc-pVTZ&	9.520 &0.0003&-0.0007&\\
$NB_2$  (X$^2\Sigma^+_g$) &cc-pVDZ& 2.566&0.0003&0.002&\\
$N_2$ (X$^2\Sigma^+_g$)&cc-pVTZ&15.444 &0.005&0.008&15.58\\
$C_2H_4$ (X$^2B_{3u}$)&cc-pVDZ&10.362&-0.023& 0.023&10.51\\
 \hline
\end{tabular}
\end{table}
Table \ref{tab:XFNO1} presents ionization energies (IPs; -- the target-reference energy gaps for the IP variant) computed using $M_1$ and the errors in the $M_2$ and $M_3$ computed IPs, relative to to $M_1$, (all reported eV).
 Except for $C_2H4$, $M_2$ mostly overestimates the IPs relative to $M_1$, whereas $M_3$ does not exhibit any such systematic trend.
For both $M_2$ as well as $M_3$, the MAE is very small -- just 0.023 eV, which is smaller than the MAE of $M_1$ relative to the experimental results.

\begin{table}[h!]
\centering
\renewcommand{\arraystretch}{1.5} 
\setlength{\tabcolsep}{4pt} 
\tiny 
\caption{\small EOM-CCSDT energies and relative errors in $M_2$ and XFNO-EOM-CCSDT energies for DIP states (in electron-volts).}
\protect\label{tab:XFNO2}
\begin{tabular}{*{7}{>{\tiny \arraybackslash}c}}
\hline
 \tiny Molecules& \tiny Basis &  \tiny $E_{\text{$M_1$}}$   &  \tiny $\delta E_{\text{$M_2$}}$ &  \tiny $  \delta E_{\text{$M_3$}}$ & $Exp$   \\  
 \\ 
\hline
\hline
\renewcommand{\arraystretch}{1.2} 
$CO$ ($^1\Sigma^+$)  &aug-cc-pVTZ& 41.4185&0.009&0.008&41.7\\
$C_2H_2$ ($^1\Delta_g$) &aug-cc-pVTZ& 33.3704&-0.003&0.002&33.0\\
$C_2H_2$ ($^1\Pi_u$) &aug-cc-pVTZ& 33.3704&-0.003&0.002&33.7\\
$HCHO$ ($^1A_1$) &aug-cc-pVDZ& 32.9062&0.007&0.002\\
$HCHO$ ($^1A_2$) &aug-cc-pVDZ& 36.6666&0.006&-0.003\\
$C_2H_4$ ($^1A_g$)&aug-cc-pVDZ& 30.706&-0.001&-0.001&30.01\\
$C_2H_4$ ($^1A_g$)&aug-cc-pVDZ&32.2573& 0.002&-0.007&32.2\\
$C_2H_4$ ($^1B_g$)&aug-cc-pVDZ&34.7161& 0.002&-0.001&34.0\\
$H_2O$ ($^1A_1$)&aug-cc-pVTZ& 41.4101&-0.004&0.003&41.3\\
$H_2O$ ($^1B_1$)&aug-cc-pVTZ& 42.8292&-0.0053&-0.001&42.0\\
 \hline
\end{tabular}
\end{table}
Table \ref{tab:XFNO2} presents the double-ionization potentials (DIPs; -- the target-reference energy gaps for the DIP variant) are reported along with relative errors for $M_2$ and $M_3$ (relative to $M_1$). The MAEs are impressively very small —- 0.009 eV for $M_2$ and 0.008 eV for $M_3$, which are smaller than the CCSDT error-bar and also smaller than the $M_1$ absolute errors relative to the experimental values.

 \subsection{Cyclobutadiene: Adiabatic TSG using DIP-EOM-CCSDT method}
 Table \ref{tab:C4H4} presents the total electronic energies ground $X ^{1}A_{g}$ and first triplet $1^{3}B_{1g}$ states and the adiabatic triplet-singlet gap (TSG) of cyclobutadiene.
 Each electronic state was computed as the DIP-EOM-CCSDT target state using the closed-shell dianomic state of the molecule as the reference computed at the target-state geometry.
 The electronic structure of cyclobutadiene is very interesting
  as its antiaromatic character provides valuable insight into the destabilizing effects of electron delocalization in small cyclic $\pi$-systems. Cyclobutadiene have the degenerate closed-shell ground state, whereas their lowest triplet state are highly reactive. We present the TEs of the ground and
the lowest triplet states, and the adiabatic TSGs (excluding the zero-point energy corrections). While $M_2$ overestimates ground and the triplet state TEs by $\sim$1.8 millihartrees, relative to $M_1$, resulting in marginal underestimation of the adiabatic TSG only by 0.03 millihartrees, $M_3$ underestimates the TEs by $\sim$0.7 millihartrees resulting in underestimation of the adiabatic TSG by only 0.01 millihartrees.

 \begin{table}[h]
\centering
\caption{Electronic Energies and Adiabatic TSGs (in Hartrees) of cyclobutadiene}
\label{tab:C4H4}
\renewcommand{\arraystretch}{1.2} 
\setlength{\tabcolsep}{1pt} 
\begin{tabular}{lcccccc}
\hline
\textbf{\tiny{States}} & 
&\textbf{\tiny{$M_1$}} & \textbf{\tiny{$M_2$}}  & \textbf{\tiny{$M_3$}}\\
 \hline
\hline
 \multirow{2}{*}{\tiny{C$_4$H$_4$}} &
\tiny $E_{S_0}(S_0=X ^{1}A_{g})$ & \tiny-154.24124 & \tiny-154.23937 		&\tiny-154.24198 \\
& \tiny $E_{T_1}(T_1 = 1^{3}B_{1g})$  & \tiny-154.22161 	 	& \tiny-154.21977 	& \tiny-154.22237 \\
&\tiny{$\Delta E_{T_1-S_0}$} & \tiny0.01962 	&\tiny0.01959  	&\tiny0.01961\\
\hline

 \end{tabular}
\end{table}

 \subsection{FNO- and XFNO-EOM-CCSDT methods for EA and DEA states}
 Table~\ref{tab:TEADEA},\ref{tab:EADEA} summarizes TEs of low-lying EA and DEA states of the molecules as well as the corresponding one- and two-electron attachment energies. A comparison between the EA/DEA and IP/DIP results reveals that the performance of the EA and DEA methods is considerably poorer. However, the poor performance of EA/DEA is not unobvious, as the electron-attachment process involves the virtual orbital subspace which is transformed and truncated in the FNO algorithm. 

 For EA, the MAEs in TEs are 5 millihartrees (0.14 eV) for $M_2$ and 3 (0.08 eV) millihartrees for $M_3$ which is of the order of the typical EA-EOM-CCSDT error bar of $\sim$0.1 eV. The performance of DEA deteriorates even more, with much larger MAEs in the TEs, that is, respectively 95 and 89 millihartrees for $M_2$ and $M_3$.  

$M_2$ and $M_3$ perform somewhat better for the (first electron-affinities, EAs), with MAEs relative to $M_1$ being respectively 0.086 eV and 0.007 eV, which are within the typical EA-EOM-CCSDT error bar ($\sim$0.1 eV).

However, the two-electron attachment energies predicted by FNO and XFNO methods are rather large with MAEs relative to $M_1$ are larger than 2 eV for both $M_2$ as well as $M_3$, which may be due to large amount of orbital relaxation effects that add to the error due to orbital-subspace-truncation.

\begin{table}[h!]
\centering
\renewcommand{\arraystretch}{1.5} 
\setlength{\tabcolsep}{4pt} 
\tiny 
\caption{\small EOM-CCSDT total energies (in hartrees) and relative errors in FNO-CCSDT ($M_2$) and XFNO-EOM-CCSDT  ($M_3$) energies for EA and DEA states (in hartrees).}
\protect\label{tab:TEADEA}
\begin{tabular}{*{7 }{>{\tiny \arraybackslash}c}}
\hline
 \tiny Molecules& \tiny Basis &  \tiny $E_{\text{$M_1$}}$   &  \tiny $\delta E_{\text{$M_2$}}$ &  \tiny $  \delta E_{\text{$M_3$}}$   \\  
 \\ 
\hline
\hline
\renewcommand{\arraystretch}{1.2} 
EA&&&&\\
$C_2$ (X$^1\Sigma_g^+$)&cc-pVTZ&-75.88681&0.005&0.003\\
$O_3$ (X$^1A_1$)&cc-pVDZ&-224.920559&0.002&0.001\\ 
 \hline
DEA&&&&\\
$CH_2^{2+}$ (X$^3B_1$)&TZ2P&-39.048606&0.095&0.089\\
$O_2^{2+}$ (X$^1\Sigma_g^+$)&cc-pVTZ&-150.123933&0.0535&0.0452\\
\hline

\end{tabular}
\end{table}

\begin{table}[h!]
\centering
\renewcommand{\arraystretch}{1.5} 
\setlength{\tabcolsep}{4pt} 
\tiny 
\caption{\small EOM-CCSDT energies and relative errors in $M_2$ and XFNO-EOM-CCSDT energies for EA and DEA states (in electron-volts).}
\protect\label{tab:EADEA}
\begin{tabular}{*{6}{>{\tiny \arraybackslash}c}}
\hline
 \tiny Molecules& \tiny Basis &  \tiny $E_{\text{$M_1$}}$   &  \tiny $\delta E_{\text{$M_2$}}$ &  \tiny $  \delta E_{\text{$M_3$}}$    \\  
 \\ 
\hline
\hline
\renewcommand{\arraystretch}{1.2} 
 EA&&&&\\
 $C_2$ (X$^1\Sigma_g^+$)&cc-pVTZ&2.880&0.0862&0.0073\\
 $O_3$ (X$^1A_1$)&cc-pVDZ&0.2993&0.0143&0.002\\
\hline
DEA&&&&\\
$CH_2^{2+}$(X$^3B_1$)&TZ2P&31.856&2.593&2.435\\
$O_2^{2+}$ (X$^1\Sigma_g^+$)&cc-pVTZ&36.797&1.356&1.253\\
\hline
\end{tabular}
\end{table}

\section{Conclusion}
We have presented FNO- and XFNO approximations for IP, DIP, EA and DEA variants of EOM-CCSDT for computing target-state total energies and the target-reference energy-gaps. 
For IP- and DIP-EOM-CCSDT, the trends in the target-state total energies exhibited by FNO and XFNO are very similar to the FNO-CCSDT and XFNO-CCSDT ground state total energies presented in our earlier work \cite{FNOCCSDT}. The FNO-EOM-CCSDT computed reference-target gaps (IPs and DIPs) are quite accurate even for lower FNO-thresholds (see the SI for more information). The FNO with $OCCT=99.9\%$ and for XFNO computed IPs and DIPs differ from the conventional EOM-CCSDT values only by 0.023 eV.

In case of EA and DEA variants, the FNO and XFNO errors are rather large compared to the conventional EOM-CCSDT values for both total energies as well as for the target-reference energy gaps. Two factors play crucial role here. Firstly, the EA and DEA-EOM-CCSDT error bars are larger than the corresponding IP and DIP counterparts, in general. Secondly, in case of FNO approach, the virtual orbital subspace is transformed and truncated which has direct impact on the EA and DEA amplitudes unlike in case of IP and DIP. Nevertheless, for EA-EOM-CCSDT, the target-reference gaps obtained by FNO- and XFNO- differ only by 0.086 eV or less from the conventional EOM-CCSDT values. This is within the EA-EOM-CCSDT error bar of 0.1 eV, although the absolute errors in the total energies of these states are slightly larger than 0.1 eV.
For DEA-EOM-CCSDT, the FNO and XFNO approaches fail badly as the EOM vectors, ($2p$, $3p-1h$ and $4p-2h$) are badly affected by truncation of the virtual subspace.

In short, for IP and DIP-EOM-CCSDT, both FNO and XFNO turn out to be very good cost-effective tools. For EA-EOM-CCSDT, it may be okay to go for FNO/XFNO to get semi-qualitative predictions about electron-affinities, but not so okay for total energies of the corresponding target-states. For DEA-EOM-CCSDT, use of FNO/XFNO techniques is strongly discouraged due to large errors in energies as well as energy-gaps.

\printcredits


\bibliographystyle{elsarticle-num}

\bibliography{cas-refs}


\end{document}